# Study of the solar wind-magnetosphere coupling on different time scales

**Badruddin and O. P. M. Aslam**
*Department of Physics, Aligarh Muslim University, Aligarh-202002, India.*
*e-mail: badr.physamu@gmail.com*

**ABSTRACT**
Solar wind-magnetosphere coupling, its causes and consequences have been studied for the last several decades. However, the assessment of continuously changing behaviour of the sun, plasma and field flows in the interplanetary space and their influence on geomagnetic activity is still a subject of intense research. Search for the best possible coupling function is also important for space weather prediction. We utilize four geomagnetic indices (*ap*, *aa*, *AE* and *Dst*) as parameters of geomagnetic activity level in the earth's magnetosphere. In addition to these indices, we utilize various solar wind plasma and field parameters for the corresponding periods. We analyse the geomagnetic activity and plasma/field parameters at yearly, half-yearly, 27-day, daily, 3-hourly, and hourly time resolutions. Regression analysis using geomagnetic and solar wind data of different time resolutions, over a continuous long period, and at different phases of solar activity (increasing including maximum/decreasing including minimum) led us to suggest that two parameters $BV/1000$ (mV m$^{-1}$) and $BV^2$ (mV s$^{-1}$) are highly correlated with the all four geomagnetic activity indices not only at any particular time scale but at different time scales. It probably suggests for some role of the fluctuations/variations in interplanetary electric potential, its spacial variation [i.e., interplanetary electric field $BV$ (mV m$^{-1}$)] and/or time variation [$BV^2$ (mV s$^{-1}$)], in influencing the reconnection rate.

## 1. Introduction

In the area of solar-terrestrial physics, one of the key problems is to investigate the mechanism of energy transfer from the solar wind into the magnetosphere. Another related key issue is to investigate the mechanism that excites magnetic disturbances in the geo-magnetosphere. It is generally believed that the basic parameter leading to geomagnetic disturbances is the southward component of the interplanetary magnetic field ($-B_z$) and/or the duskward component of the interplanetary electric field $E_y = -V\text{x}B_z$ (see, e.g., Dungey, 1961; Rostoker and Fälthammar, 1967; Burton et al., 1975; Akasofu, 1981; Badruddin, 1998; Echer et al., 2005; Gopalswamy et al., 2008; Badruddin and Singh, 2009; Alves et al., 2011; Guo et al., 2011; Singh and Badruddin, 2012; Yermolaev et al., 2012 and references therein). With negative $B_z$, reconnection occurs at the daytime magnetopause between the Earth's magnetic field and southward $B_z$ component of the interplanetary magnetic field (Kane, 2010). The principal manifestation of geomagnetic storms, measured by the index *Dst*, is the increase of ring current intensity, which depends upon the reconnection rate.

Origin of the intense southward magnetic fields are the interplanetary shock/sheath region, coronal mass ejections/magnetic clouds, stream interaction regions etc. (e.g., Lepping et al., 1991; Märcz, 1992; Tsurutani and Gonzalez, 1997; Richardson et al., 2000; Webb et al., 2000; Kudela and Storini, 2005; Kim et al., 2005; Gopalswamy et al., 2007; Singh and Badruddin, 2007; Zhang et al., 2007; Gupta and Badruddin, 2009; Yermolaev et al., 2010; Alves et al., 2011; Mustajab and Badruddin, 2011; Richardson and Cane, 2011; Kudela, 2013) and arrival of these structures leads to changes/fluctuations in various interplanetary plasma and field parameters.

In spite of the success of the so called Dungey mechanism (Arnoldy, 1971; Burton et al., 1975; Holzer and Slavin, 1979; Alves et al., 2011) some effort (e.g., Baker et al., 1981; Clauer et al., 1981; Holzer and Slavin, 1982; Murayama, 1982; Zhang and Burlaga, 1988; Papitashvili et al., 2000; Sabbah, 2000; Gupta and Badruddin, 2009; Dwivedi et al., 2009; Joshi et al., 2011) has gone into looking for other parameters that might correlate better with geomagnetic activity.
Geomagnetic activity being influenced by total interplanetary electric field (Papitashvili et al., 2000; Sabbah, 2000), irregularities in the solar wind and interplanetary magnetic field (Dessler and Fejer, 1963; Garrett et al., 1974; Crooker et al., 1977; Kershengolts et al., 2007), and enhanced dynamic pressure (Murayama, 1982; Srivastava and Venkatakrishnan, 2002; Boudouridis et al., 2005; Xie et al., 2008; Ontiveros and Gonzalez-Esparza, 2010; Singh and Badruddin, 2012) have been suggested. However, a unique relationship is still lacking which may ultimately lead to understand the intensity of geomagnetic disturbances.

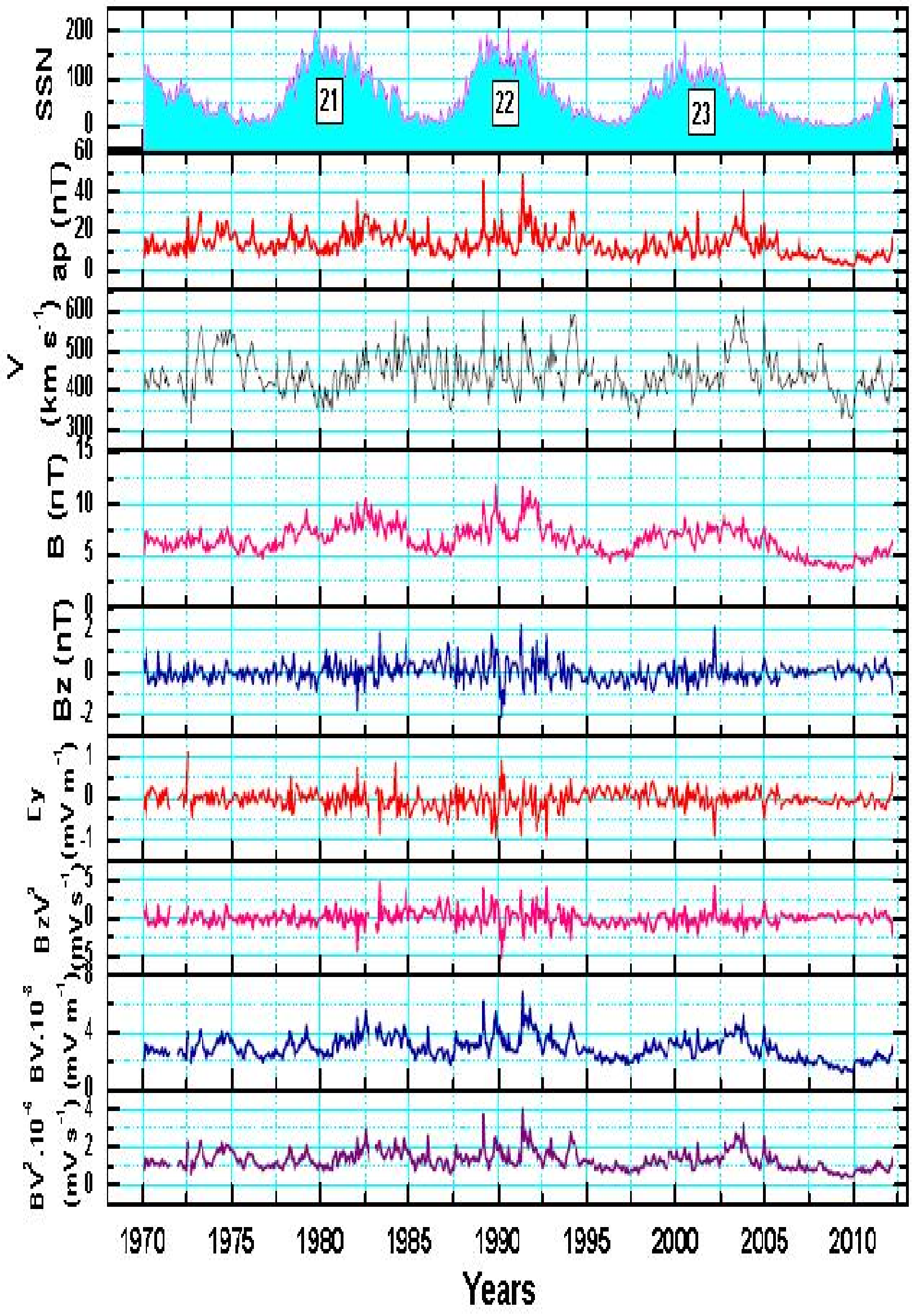

**Fig. 1.** Time variation of 27-day average solar (SSN), geomagnetic (*ap*), interplanetary plasma/field parameters *V* (km $s^{-1}$), *B* (nT), *Bz* (nT), *Ey* (mV $m^{-1}$), *BV* (mV $m^{-1}$), $BzV^2$ (mV $s^{-1}$) and $BV^2$ (mV $s^{-1}$).

Most of the earlier efforts to search for better coupling functions, in general used one geomagnetic index or the other, at one time resolution or the other. Further, earlier studies were mainly focused over the durations of moderate to strong geomagnetic disturbances. To the best of our knowledge, none of the previous correlative studies were done over extended periods of several solar cycles using different geomagnetic activity and solar wind parameters at multiple time resolutions. In this work we analyse the continuous data for long periods (~40 years), that contain quiet, weak, moderate as well as strong geomagnetic activity periods. In this paper, we present the results of the analysis using interplanetary plasma and field data and their various derivatives together with various geomagnetic indices of different time resolutions; yearly, half-yearly, 27-day, daily, 3-hourly and hourly resolutions.

## 2. Results and Discussion

In Fig. 1, we have plotted the time variation of 27-day average solar, geomagnetic and interplanetary parameters (http://omniweb.gsfc.nasa.gov) for more than three solar cycles (1970-2011). The parameters plotted in this figure are; sunspot number (SSN) -a solar activity parameter; *ap* index -a parameter of geomagnetic activity; interplanetary plasma and field parameters -solar wind velocity $V$ (km $s^{-1}$), interplanetary magnetic field $B$ (nT), its north-south component $B_z$ (nT), duskward electric field $E_y$ (mV $m^{-1}$), 'spacial variation of interplanetary electric potential' i.e., the interplanetary electric field $BV.10^{-3}$ (mV $m^{-1}$), and two more derivatives that may be referred as 'time variation of the duskward electric potential' $BzV^2$ (mV $s^{-1}$), and the 'time variation of total interplanetary electric potential' $BV^2$ (mV $s^{-1}$); although suitability of these latter nomenclatures need to be confirmed. From Fig. 1 alone, it is difficult to infer about interplanetary plasma/field parameter whose time variation best matches with time variation in geomagnetic activity level, it looks, however, as if the time variation of $BV^2$ is relatively better related to *ap* variations at this (27-day average) time resolution.

In order to understand the response of magnetosphere to varying interplanetary conditions, attempts have been made in the past to search for the parameter(s) that can best explain the occurrence of geomagnetic disturbances, but, efforts are needed to find a relationship that may ultimately lead to unambiguously understand the solar wind-magnetosphere coupling and disturbances in the geo-magnetosphere.

As solar polarity reverses at/near each solar activity maximum, we have divided a complete solar cycle into two parts; (i) increasing including maximum and (ii) decreasing including minimum phases. This division is aimed to look for, if any, the large-scale interplanetary magnetic field (IMF) polarity dependent effects of solar plasma/field parameters on the geomagnetic activity. It is to be mentioned here that large scale IMF–polarity is positive (outward above the heliospheric current sheet and inward below the heliospheric current sheet) during decreasing including minimum phases and negative (inward above the heliospheric current sheet and outward below the heliospheric current sheet) during increasing including maximum phases of even solar cycles (e.g., solar cycles 20 and 22), opposite will be the polarity during similar phases of odd solar cycles (e.g., solar cycles 21 and 23).

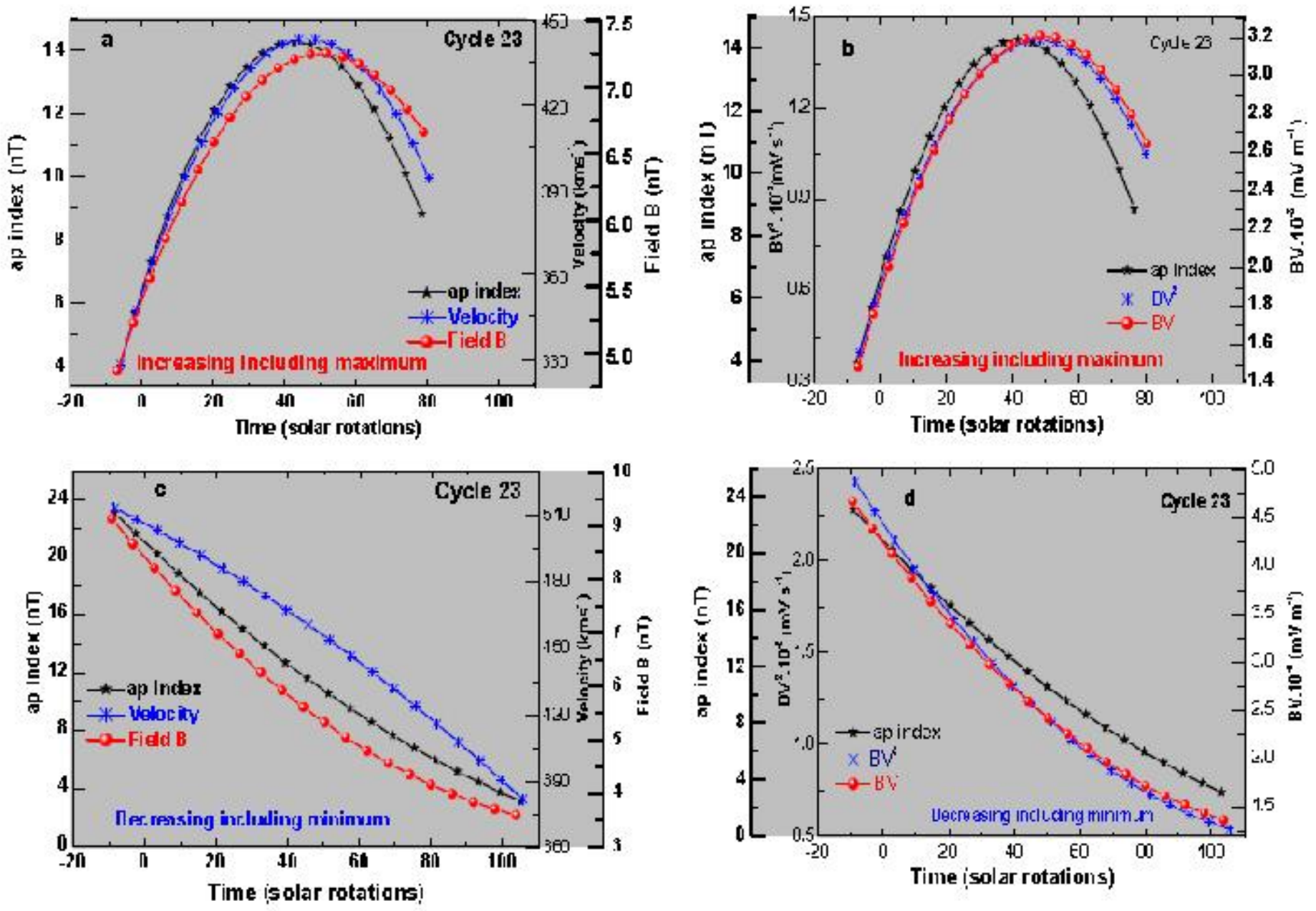

**Fig. 2.** Best-fit polynomial curves ($y = A+B_1x+B_2x^2$) for different interplanetary parameters and geomagnetic parameter *ap* during (a and b) increasing including maximum and (c and d) decreasing including minimum of solar cycle 23.

We have adopted two approaches, (a) best-fit approach and (b) correlative analysis approach. First, we did the polynomial ($y = A+B_1x+B_2x^2$) fit to the 27-day averaged parameters (*ap*, $B$, $B_z$, $\sigma_B$, $V$, $P$, $E_y$, $BV$, $BzV^2$ and $BV^2$), both during (i) increasing including maximum phase and (ii) decreasing including minimum phase of solar cycle 23. The constants $B_1$, $B_2$ and determination coefficient ($R^2$) for the best-fit curves to the time evolutions of different parameters during both the phases of cycle 23 are given in Table 1. We can see from the values of $R^2$ that the polynomial fit is not good for all parameters. It is relatively better in case of interplanetary parameters $B$, $V$, $BV$ and $BV^2$. In order to see which of these parameters better follows the polynomial fitted curve of geomagnetic parameter *ap*, we have plotted them in Fig. 2. From these fitted curves, it is difficult to make any clear distinction between various plasma/field parameters (*B*, *V*, *BV* and $BV^2$) as regards to which of them tracks relatively better the time evolution of geomagnetic activity parameter *ap*.

**Table 1** Best fit parameters $B_1$ and $B_2$ and determination coefficient ($R^2$) obtained by fitting polynomial ($y = A + B_1x + B_2x^2$) to the time evolution of different parameters (27-day average data) during different phases solar cycle 23.

| Parameters | Decreasing including minimum phase | | | Increasing including maximum phase | | |
|---|---|---|---|---|---|---|
| | $B_1$ | $B_2$ | $R^2$ | $B_1$ | $B_2$ | $R^2$ |
| *ap* index (nT) | -0.24±0.07 | $(6.2\pm6.7) \times 10^{-4}$ | 0.55 | 0.36±0.10 | -0.004± 0.001 | 0.19 |
| *B* (nT) | -0.08±0.002 | $(2.8\pm0.78) \times 10^{-4}$ | 0.87 | 0.07±0.02 | $-(7.2\pm2.0) \times 10^{-4}$ | 0.38 |
| $B_z$ (nT) | 0.002±0.004 | $(6.3\pm4.2) \times 10^{-4}$ | 0.08 | 0.006±0.01 | $(1.8\pm1.5) \times 10^{-4}$ | 0.08 |
| Sigma B (nT) | -0.002±0.01 | $-(1.8\pm0.97) \times 10^{-4}$ | 0.42 | 0.03±0.02 | $-(2.5\pm2.2) \times 10^{-4}$ | 0.09 |
| *V* (km $s^{-1}$) | -0.87±0.7 | -0.003±0.007 | 0.32 | 3.87±0.66 | -0.04±0.009 | 0.39 |
| *P* (n Pa) | -8.65±0.005 | $-(1.2\pm4.8) \times 10^{-4}$ | 0.54 | -0.01±0.007 | $(8.8\pm9.7) \times 10^{-5}$ | 0.02 |
| $E_y$(mV $m^{-1}$) | $-4.1\times10^{-4}\pm0.002$ | $(1.02\pm1.95) \times 10^{-5}$ | 0.09 | -0.002±0.005 | $-(2.1\pm6.3) \times 10^{-5}$ | 0.09 |
| $B_zV^2 .10^{-5}$ (mV $s^{-1}$) | -0.007±0.01 | $-(5.14\pm9.9) \times 10^{-6}$ | 0.07 | -0.003±0.02 | $(1.66\pm2.72) \times 10^{-4}$ | 0.05 |
| $BV^2 .10^{-6}$ (mV $s^{-1}$) | -0.03±0.005 | $-(9.88\pm5.3) \times 10^{-5}$ | 0.63 | 0.034±0.006 | $-(3.54\pm0.74) \times 10^{-5}$ | 0.43 |
| $BV.10^{-3}$ (mV $m^{-1}$) | -0.04±0.007 | $-(1.57\pm0.67) \times 10^{-4}$ | 0.77 | 0.06±0.009 | $-(5.7\pm1.2) \times 10^{-4}$ | 0.45 |

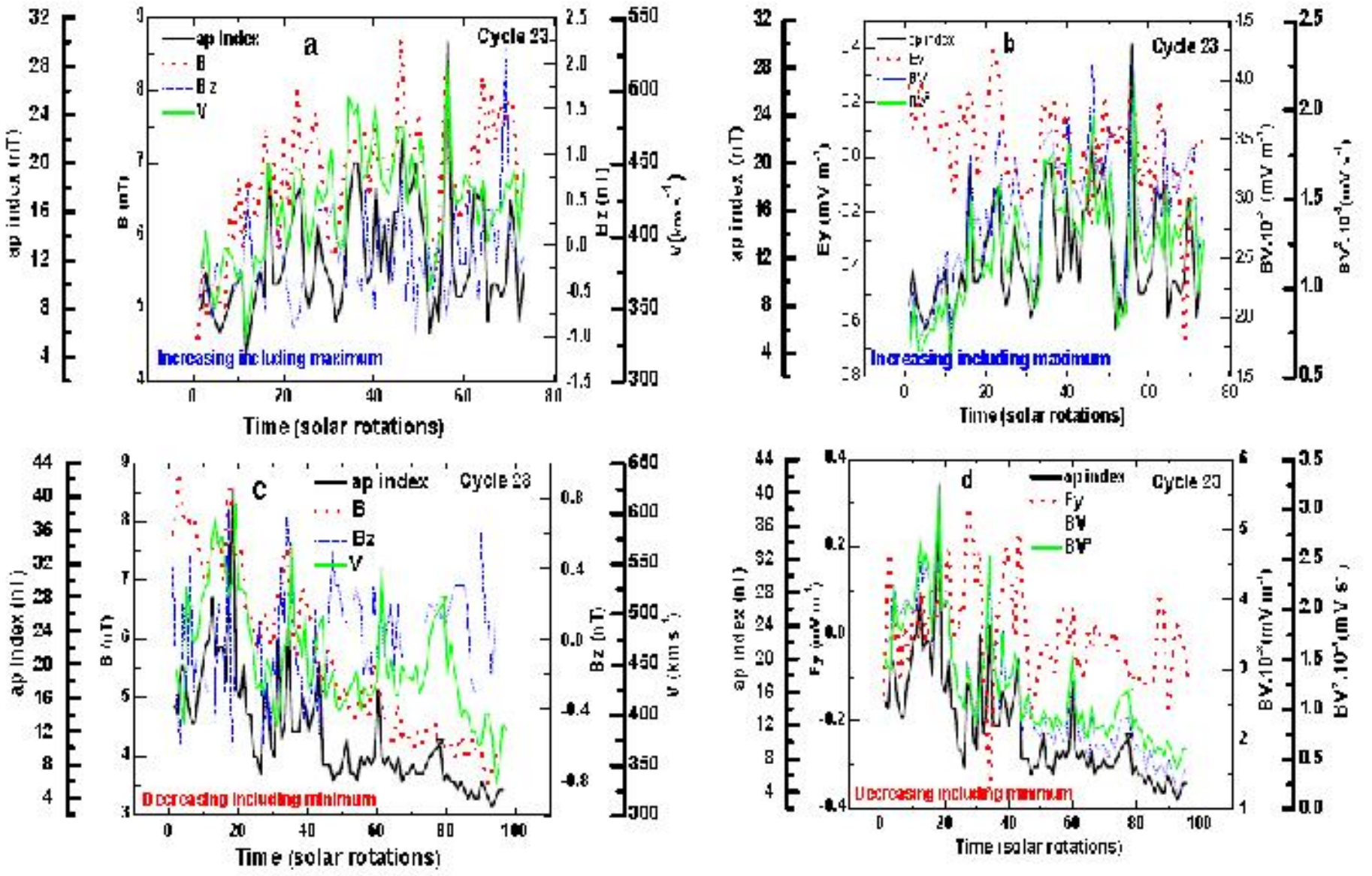

**Fig. 3**. Time variation (27-day solar rotation average) of various interplanetary plasma/field parameters with geomagnetic *ap* index during increasing including maximum (a and b) and decreasing including minimum (c and d) phases of solar cycle 23.

Therefore, as a next step, we have considered the time variations in different parameters at this time scale (27-day), both during increasing including maximum and decreasing including minimum phases of solar cycle 23. These time variations in various interplanetary plasma/field parameters (*V*, *B*, $B_z$, $E_y$, *BV* and $BV^2$) are compared with time variations of geomagnetic parameter *ap* (see Fig. 3 (a-d)). From these figures, it appears that, at this time scale, *BV* and $BV^2$ follow better the time variation of *ap*, as compared to other parameters *V*, *B*, $B_z$ and $E_y$ considered here.

**Table 2a**
**Rate of change of the *ap* index with various parameters (Δ*I*/Δ*P*) and correlation coefficient (*R*) during increasing including maximum phases of solar cycles 20, 21, 22 and 23 using 27-day average data.**

| Parameters | Solar cycle 20 | | Solar cycle 21 | | Solar cycle 22 | | Solar cycle 23 | |
|---|---|---|---|---|---|---|---|---|
| | Δ*I*/Δ*P* | *R* | Δ*I*/Δ*P* | *R* | Δ*I*/Δ*P* | *R* | Δ*I*/Δ*P* | *R* |
| *B* (nT) | 3.44±0.67 | 0.57 | 2.81±0.54 | 0.55 | 3.99±0.45 | 0.74 | 3.77±0.48 | 0.68 |
| *Bz* (nT) | -3.44±1.0 | -0.42 | -1.32±1.34 | -0.13 | -3.08±1.35 | -0.28 | -2.65±1.09 | -0.29 |
| Sigma B (nT) | 3.3±0.86 | 0.46 | 3.08±0.80 | 0.44 | 4.87±0.76 | 0.62 | 4.17±0.53 | 0.68 |
| *V* (km $s^{-1}$) | 0.08±0.01 | 0.69 | 0.10±0.01 | 0.78 | 0.12±0.01 | 0.77 | 0.10±0.01 | 0.77 |
| *P* (n Pa) | 8.08±1.35 | 0.35 | 5.41±1.06 | 0.54 | 7.43±0.67 | 0.80 | 7.62±1.43 | 0.53 |
| *Ey*(mV $m^{-1}$) | 9.38±1.62 | 0.61 | 2.3±2.82 | 0.10 | 5.76±2.93 | 0.24 | 6.27±2.46 | 0.29 |
| $BzV^2.10^{-5}$(mV $s^{-1}$) | -1.80±0.52 | -0.42 | -0.83±0.73 | -0.14 | -1.34±0.63 | -0.25 | -1.72±0.56 | -0.34 |
| $BV^2.10^{-6}$ (mV $s^{-1}$) | 11.4± 1.32 | 0.77 | 11.4± 1.02 | 0.82 | 18.53±0.72 | 0.90 | 12.01±0.93 | 0.84 |
| $BV.10^{-3}$ (mV $m^{-1}$) | 7.3±0.84 | 0.76 | 6.35±0.68 | 0.76 | 7.38±0.51 | 0.87 | 7.0±0.60 | 0.81 |

The plots in Fig. 3 provide only a qualitative idea about the time variations of various plasma/field parameters as compared to geomagnetic parameter *ap* during (a) increasing including maximum and (b) decreasing including minimum phase of solar cycle 23. Therefore, a quantitative analysis has been done by best-fit linear regression analysis, not only during (a) increasing including maximum and (b) decreasing including minimum phase of solar cycle 23, and but also during similar phases of solar cycles 20, 21 and 22. The rate of change of the *ap* index with various plasma/field parameters (Δ*I*/Δ*P*) obtained from the best-fit method and linear correlation coefficients between them are tabulated in Table 2a during increasing including maximum phases of solar cycles 20, 21, 22 and 23. From this table we observe that, out of various plasma/field parameters, the correlations are found to be best with parameters *BV* (mV $m^{-1}$) or $BV^2$ (mV $s^{-1}$), consistently during increasing including maximum phases of all four solar cycles 20, 21, 22 and 23, at this time resolution (27-day) of the data. The scatter plot and best-fit linear curves of *ap* with *BV* and $BV^2$ are shown in Fig. 4a and Fig. 4c.

**Table 2b**
**Rate of change of the *ap* index with various parameters (Δ*I*/Δ*P*) and correlation coefficient (*R*) during decreasing including minimum phases of solar cycles 20, 21, 22 and 23 using 27-day average data.**

| Parameters | Solar cycle 20 | | Solar cycle 21 | | Solar cycle 22 | | Solar cycle 23 | |
|---|---|---|---|---|---|---|---|---|
| | Δ*I*/Δ*P* | *R* | Δ*I*/Δ*P* | *R* | Δ*I*/Δ*P* | *R* | Δ*I*/Δ*P* | *R* |
| *B* (nT) | 4.20±0.69 | 0.54 | 3.08±0.33 | 0.76 | 2.62±0.42 | 0.60 | 3.76±0.26 | 0.83 |
| *Bz* (nT) | 2.10±1.69 | 0.14 | -3.85±1.29 | -0.37 | 0.03±1.21 | 0.003 | -6.47±2.20 | -0.31 |
| Sigma B (nT) | 3.6±0.77 | 0.44 | 4.43±0.69 | 0.63 | 3.02±0.58 | 0.54 | 5.86±0.53 | 0.75 |
| *V* (km $s^{-1}$) | 0.07±0.01 | 0.78 | 0.07±0.01 | 0.53 | 0.06±0.009 | 0.64 | 0.09±0.007 | 0.81 |
| *P* (n Pa) | 6.83±1.16 | 0.55 | 4.89±1.22 | 0.46 | 6.99±1.01 | 0.65 | 10.7±1.89 | 0.77 |
| *Ey* (mV $m^{-1}$) | -0.73±3.72 | -0.02 | 7.2±2.56 | 0.34 | 0.56±2.67 | 0.03 | 14.8±4.85 | 0.30 |
| $BzV^2.10^{-5}$ (mV $s^{-1}$) | 1.22±0.81 | 0.16 | -1.25±0.59 | -0.26 | 0.19±0.58 | 0.04 | -2.77±0.96 | -0.29 |
| $BV^2.10^{-6}$ (mV $s^{-1}$) | 11.2±0.72 | 0.85 | 10.8± 0.84 | 0.86 | 94.2±0.89 | 0.79 | 11.31±0.44 | 0.93 |
| $BV.10^{-3}$ (mV $m^{-1}$) | 7.8±0.58 | 0.83 | 6.51±0.48 | 0.87 | 5.98±0.59 | 0.78 | 6.5±0.30 | 0.92 |

Similar correlative analysis during decreasing including minimum phases of solar cycles 20, 21, 22 and 23 were also done; the values of the rate of change of geomagnetic index *ap* with various parameters (Δ*I*/Δ*P*) and correlation coefficients are tabulated in Table 2b. From Table 2b, we observe that out of all parameters considered here, the *BV* and $BV^2$ are correlated better with the *ap* during decreasing including minimum phases of almost all the four cycles 20, 21, 22 and 23 averaged over the solar rotation time scale. The scatter plots with best-fit curve during decreasing including minimum phase of all four cycles are shown in Fig. 4b and Fig. 4d. A comparison of the values of $\Delta I/\Delta P$ and *R* in Table 2a and Table 2b does not show any large-scale IMF-polarity dependent change in these values.

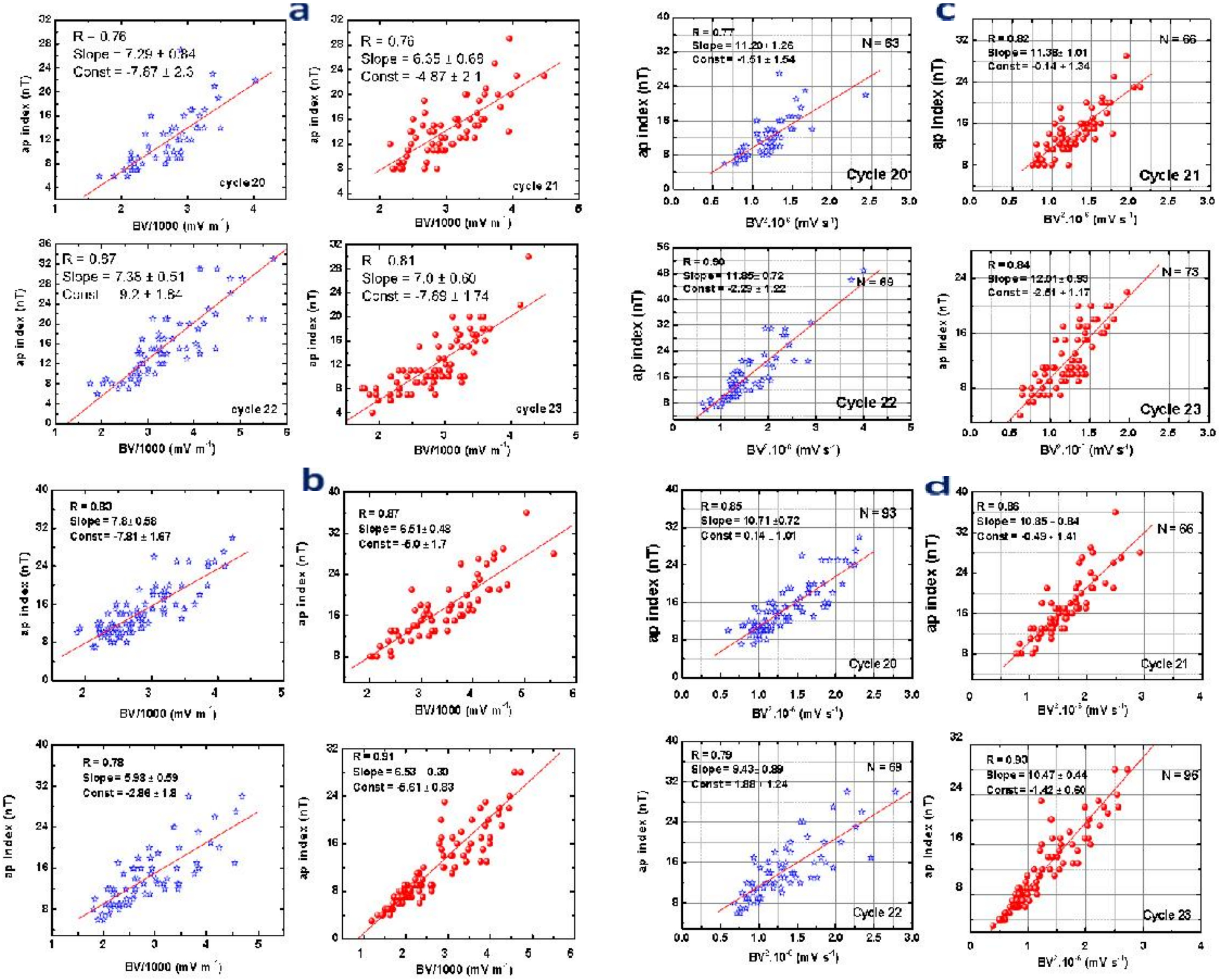

**Fig. 4.** (a) Scatter plot with best-fit linear curve between *ap* and *BV* during increasing including maximum period of different solar cycles (20 – 23). (b) Scatter plot with the best-fit linear curve between *ap* and *BV* during decreasing including minimum period of different solar cycles (20 – 23). (c) Scatter plot with best-fit linear curve between *ap* and $BV^2$ during increasing including maximum period of different solar cycles (20 – 23). (d) Scatter plot with the best-fit linear curve between *ap* and $BV^2$ during decreasing including minimum period of different solar cycles (20 – 23).

As we do not see any major difference in the values of $\Delta I/\Delta P$ during different solar cycles, as a next step, to find the best relation between geomagnetic indices and plasma/field parameters, we have combined the data of all four solar cycles during (a) increasing including maximum and (b) decreasing including minimum phases of solar cycles and did regression analysis of the combined data. To see the consistency we did a similar analysis with another geomagnetic index (*aa*) (http://ftp.ngdc.noaa.gov) also. The scatter plots of geomagnetic parameters (*aa* and *ap*) with *BV* and $BV^2$ for (i) increasing including maximum phase [Fig. 5 (a and b), Fig. 6 (a and b)], (ii) decreasing including minimum phase [Fig. 5 (c and d), Fig. 6 (c and d)] and (iii) combined periods [Fig. 5 (e and f), Fig. 6 (e and f)] with values of $\Delta I/\Delta P$ and $R$ are shown in Fig. 5 and Fig. 6. From these figures, we can give best-fit relationship for solar rotation averaged data as follows:

$$ap = 6.72x10^{-3}\,(BV) - 5.94$$
$$ap = 1.10x10^{-5}\,(BV^2) - 0.83$$
$$aa = 8.77x10^{-3}\,(BV) - 2.87$$
$$aa = 1.45x10^{-5}\,(BV^2) + 3.56$$

We have analysed and discussed the results based on the solar-rotation (27-day) averaged data. However, as shown in Fig. 7 (a, b and c), there are distinct fluctuations in the time variability of parameters at different time averages; 27-day solar rotation average [Fig. 7 (a)], 1 day (daily) average [Fig. 7 (b)] and 1 hour (hourly) average [Fig. 7 (c)] data, plotted for solar cycle 23. We, therefore, performed correlation analyses using not only the solar cycle 23 data (most complete data as compared to previous cycles 20, 21 and 22) but also data for extended period (1970-2011) at various (yearly, half-yearly, 27-day, daily, 3-hourly and hourly) time resolutions. The calculated values of linear correlation coefficients ($R$) between geomagnetic parameter *ap* and various plasma/field parameters are given in Table 3a.

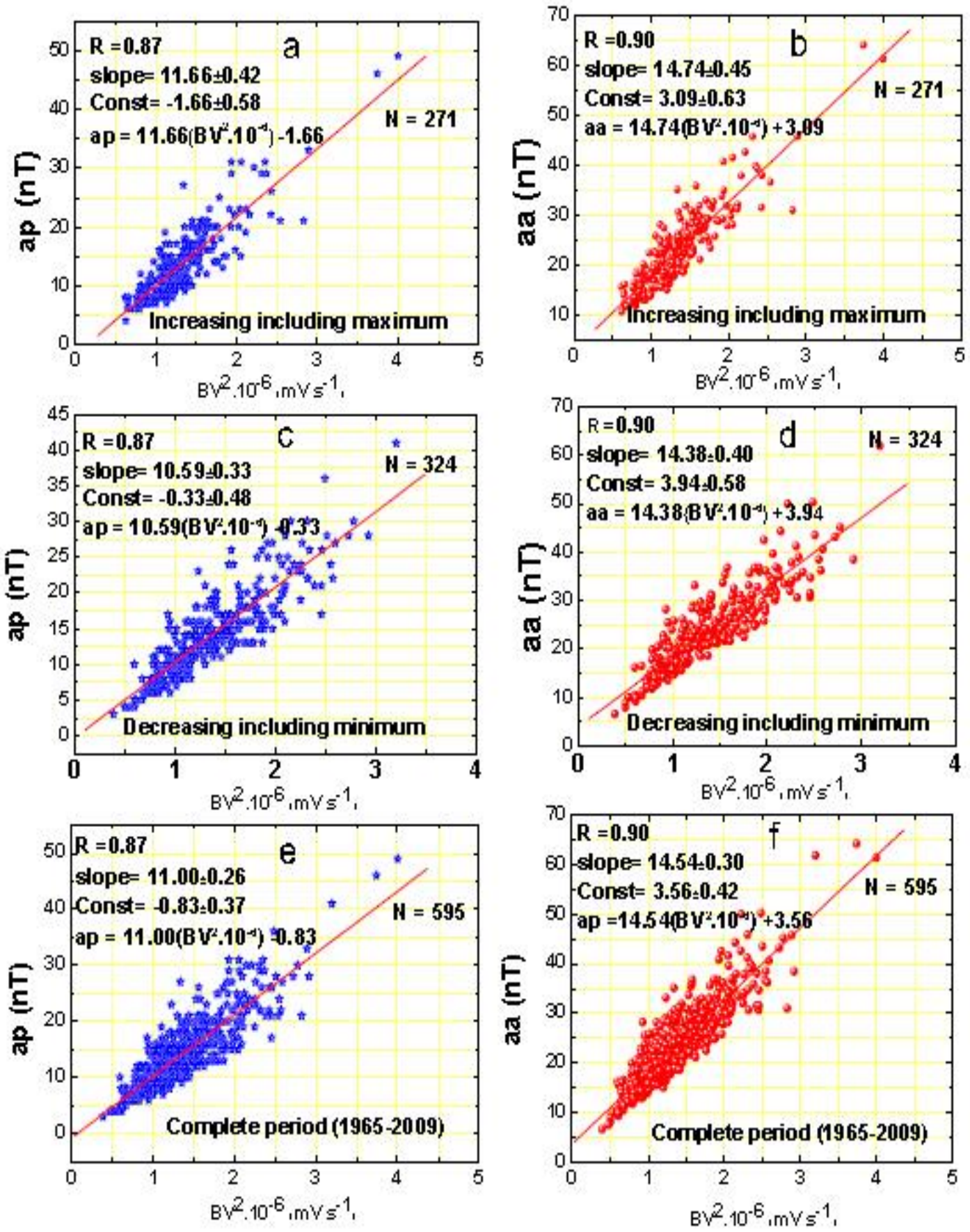

**Fig. 5.** Scatter plot and the best-fit linear curve between *ap*, *aa* and *BV* during (i) increasing including maximum phases (a and b), (ii) decreasing including minimum (c and d) phases and (iii) complete cycles (e and f) 20-23 combined.

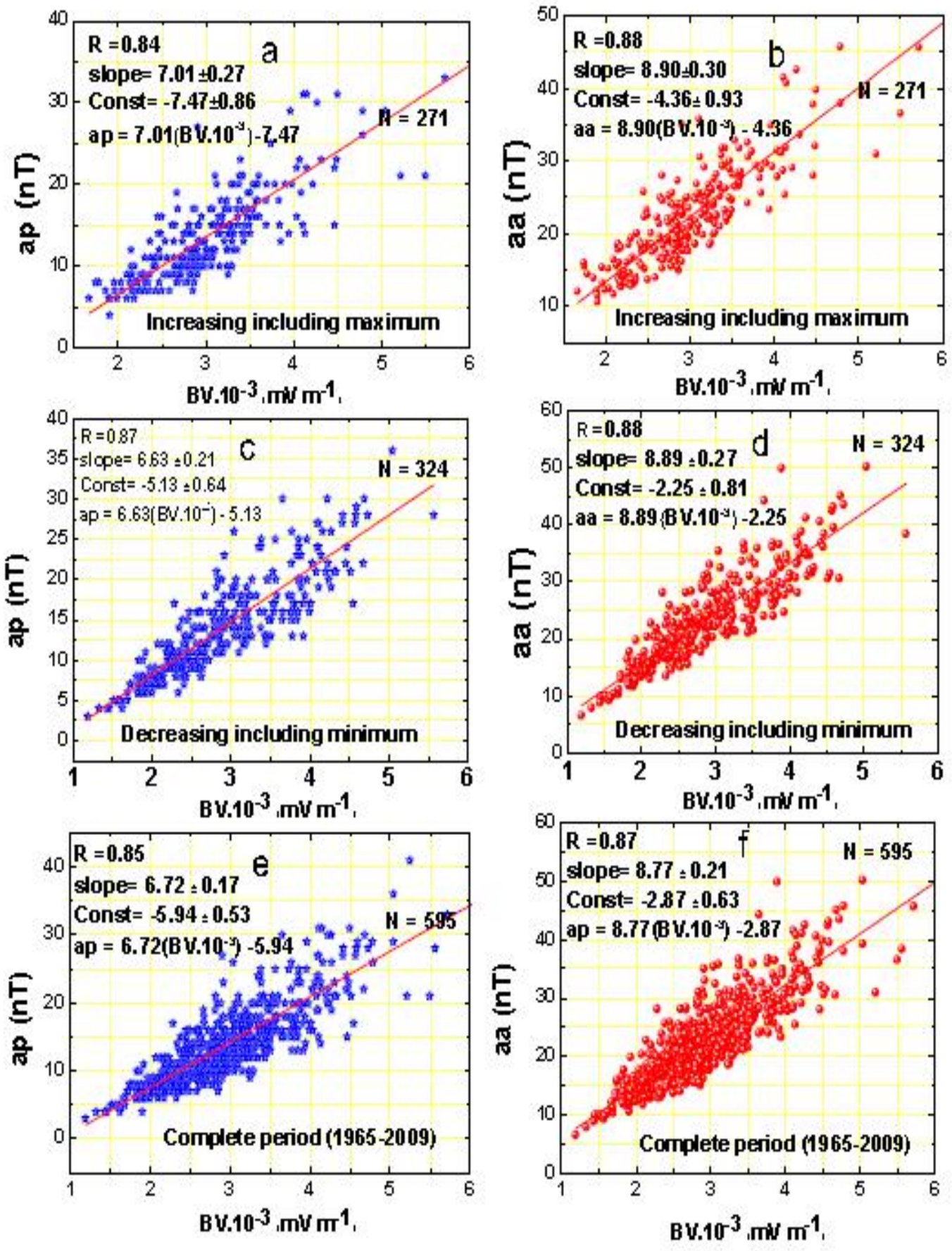

Fig. 6. Scatter plot and best-fit linear curve between *ap*, *aa* and $BV^2$ during (i) increasing including maximum phases (a and b), (ii) decreasing including minimum (c and d) phases and (iii) complete cycles (e and f) 20-23 combined

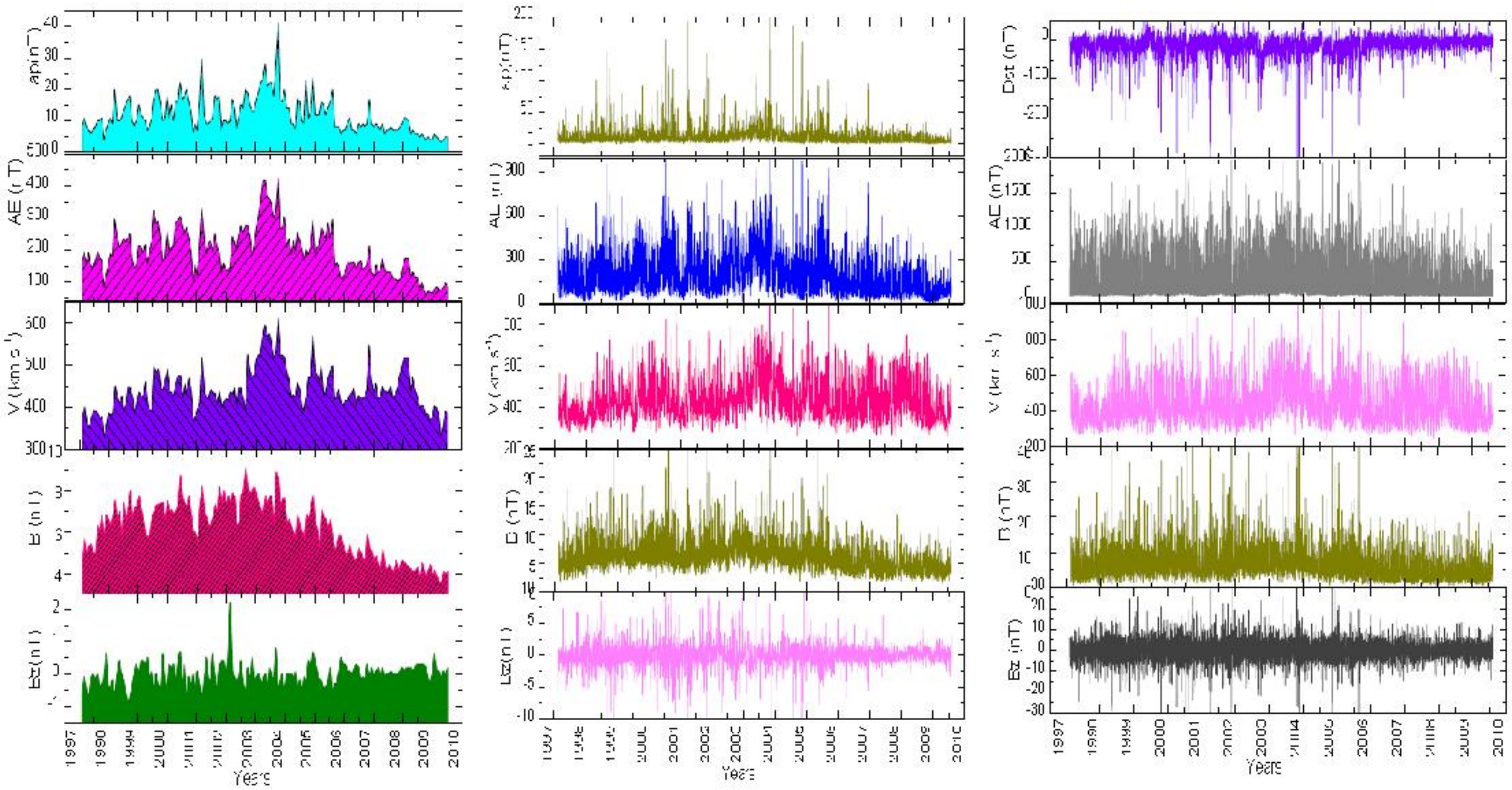

**Fig. 7.** (a) Time variation (27-day average) of various parameters during solar cycle23. (b) Time variation (daily values) of various parameters during solar cycle 23. (c) Time variation (hourly values) of various parameters during solar cycle 23.

From this table we can see that, in general, for almost all the time resolutions of data (yearly, half-yearly, 27-day, daily and 3-hourly), *BV* and $BV^2$ are highly correlated with geomagnetic parameter *ap*. We have further extended our correlation analysis at various time scales to some other geomagnetic parameters; *aa*, *AE* and *Dst* (see Table 3 (b-d)). Again, we find that *aa*, *AE* and *Dst* are also better correlated with *BV* and $BV^2$ at almost all time resolutions (yearly, half-yearly, 27-day, daily, 3-hourly and hourly). Looking at the linear correlation coefficients (*R*) in Table 3a-d at various time resolutions, we see that the values of *R* for geomagnetic indices versus $B_z$ (and also with $E_y$) are consistent at daily, 3-hourly and hourly time resolutions for all four periods considered; (1) increasing including maximum (2) decreasing including minimum (3) complete solar cycle 23 and (4) extended period 1970-2011. However, the value of *R* is of different sign (positive/negative) during increasing including maximum as compared to decreasing including minimum periods, at yearly and half-yearly time resolutions. This difference probably arises due to longer time averaging the parameters ($-B_z$ and $E_y$) which have both positive and negative values.

The relationships of geomagnetic parameters with *BV* and $BV^2$ are shown in Fig. 8 (a-d) and Fig. 9 (a-d) at various time resolutions. The equations obtained by the best-fit method are summarized in Table 4.

It is known that the primary interplanetary cause of moderate and intense geomagnetic storms is the presence of a southward interplanetary field structure in the solar wind and duskward electric field (*E*y) is a crucial parameter. However, some other 'preconditions' playing an additional role in 'influencing' the magnetic reconnection rate and further enhancing the energy transfer from the solar wind to the magnetosphere has also been studied by earlier workers.

In early, 1960's Snyder et al. (1963) showed that the geomagnetic activity *K*p responds well to solar wind velocities (*V*). Rossberg (1989) used continues data of about six months to study the effect of solar wind velocity on substorm activity using the *AL* index. He observed that at high solar wind velocities, the magnetic activity starts to increase already in the positive *B*z range and that this result poses strong constraints on the generally accepted reconnection/merging model.

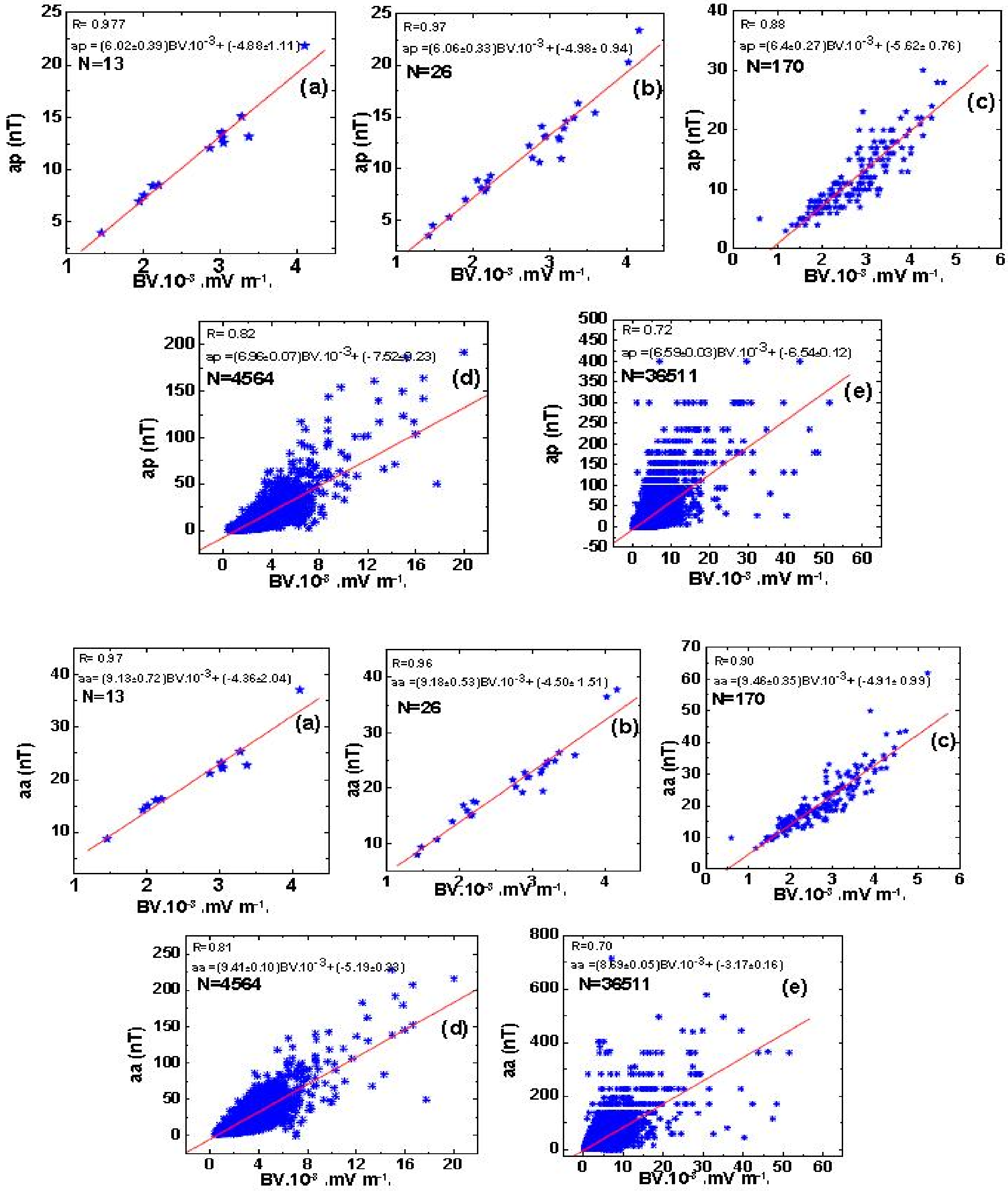

**Fig. 8.** (A) Scatter plot and best-fit linear curve between geomagnetic index ap and BV at different time resolutions; (a) yearly, (b) half-yearly, (c) 27-day, (d) daily and (e) 3-hourly. (B) Scatter plot and best-fit linear curve between geomagnetic index aa and BV at different time resolutions; (a) yearly, (b) half-yearly, (c) 27-day, (d) daily and (e) 3-hourly. (C) Scatter plot and best-fit linear curve between geomagnetic index AE and BV at different time resolutions; (a) yearly, (b) half-yearly, (c) 27-day, (d) daily, (e) 3-hourly and (f) hourly. (D) Scatter plot and best-fit linear curve between geomagnetic index Dst and BV at different time resolutions; (a) yearly, (b) half-yearly, (c) 27-day, (d) daily,(e) 3-hourly and (f) hourly.

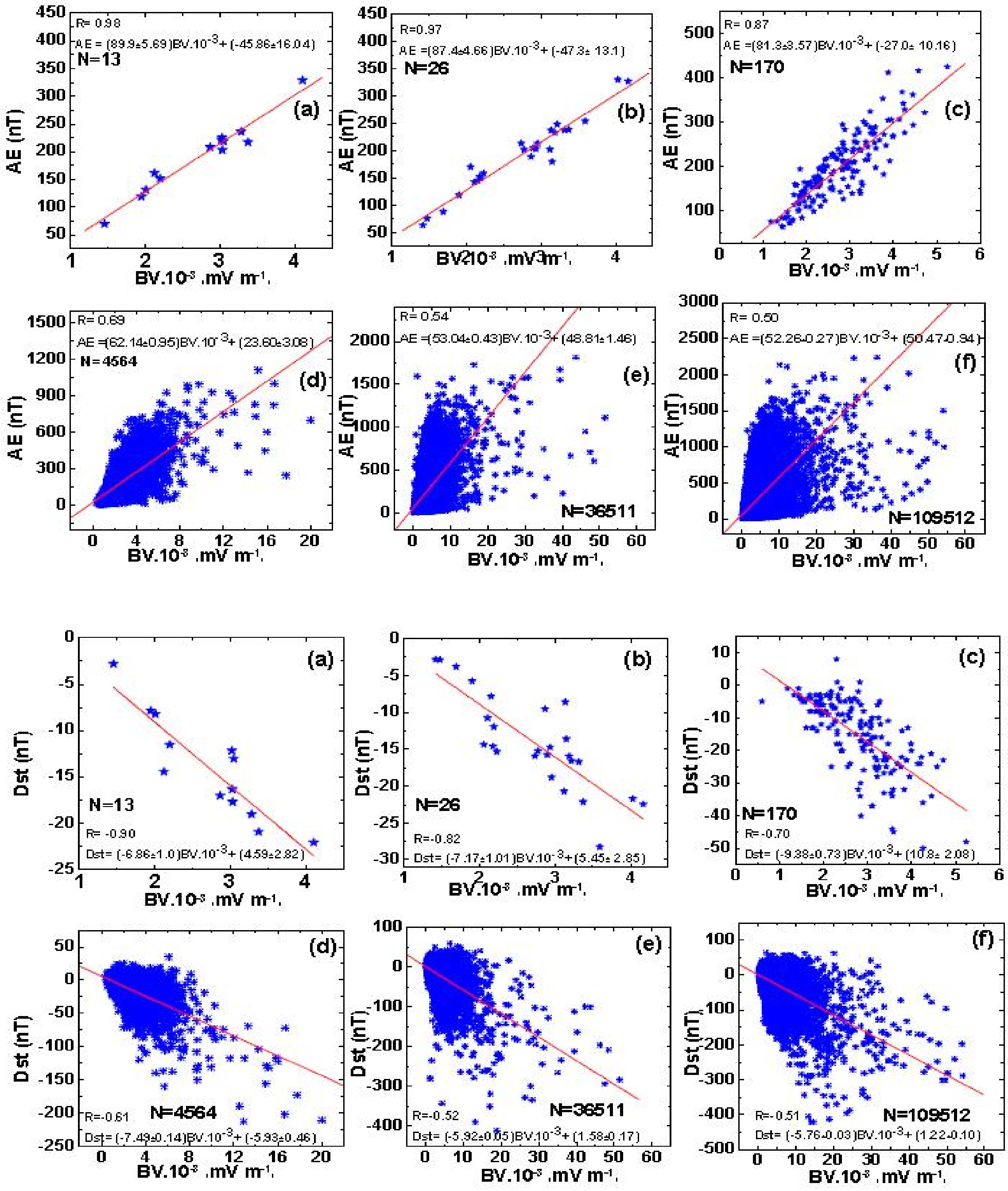

**Fig. 8.** (continued)

On yearly average scale, geomagnetic index *ap* was reported to be better correlated with $V^2$ than with $B$ (Crooker et al., 1977). However, at shorter time scales $E$y ($B$s$V$) or $B$s$V^2$ were suggested as preferred parameters (Burton et al., 1975; Murayama and Hakamada, 1975; Dessler and Hill, 1977). Garrett et al. (1974) suggested that on short time scales (~1 hour) the level of geomagnetic activity depends primarily on dawn-dusk electric field ($B$s$V$), which is proportional to $V$, but it is also enhanced by positive time derivative of solar wind velocity.

In addition to $B$s$V$ (-$B$z$V$), the products $BV$, $BV^2$, $B^2V$, $B^2$s$V$, $B$s$V$, $B$s$V^2$ were considered for evaluation as predictors of geomagnetic activity as measured by different geomagnetic parameters (Baker et al., 1981; Clauer et al., 1981; Holzer and Slavin, 1982). They preferred one or the other, and a physical meaning too remains unclear.

**Table 3a**

Linear correlation coefficient (*R*) of *ap* index with interplanetary plasma/field parameters during different phases of solar cycle 23 at different time resolutions, and for extended period (1970-2011). Best (pink) and second best (yellow) values of *R* are highlighted for each set.

| | Correlation coefficients (R) | | | | | | | | | | | | | | | | | | | | | | | |
|---|---|---|---|---|---|---|---|---|---|---|---|---|---|---|---|---|---|---|---|---|---|---|---|---|
| | Yearly | | | | | | | | Half yearly | | | | | | | | 27 days | | | | | | | |
| Periods | No. of data points | B nT | Bz nT | V $kms^{-1}$ | Ey $mVm^{-1}$ | $BV^2$ $\cdot 10^{-6}$ $mVs^{-1}$ | $BzV^2$ $\cdot 10^{-6}$ $mVs^{-1}$ | BV $.10^{-3}$ $mVm^{-1}$ | No. of data points | B nT | Bz nT | V $kms^{-1}$ | Ey $mVm^{-1}$ | $BV^2$ $\cdot 10^{-6}$ $mVs^{-1}$ | $BzV^2$ $\cdot 10^{-6}$ $mVs^{-1}$ | BV $.10^{-3}$ $mVm^{-1}$ | No. of data point | B nT | Bz nT | V $kms^{-1}$ | Ey $mVm^{-1}$ | $BV^2$ $\cdot 10^{-6}$ $mVs^{-1}$ | $BzV^2$ $\cdot 10^{-6}$ $mVs^{-1}$ | BV $.10^{-3}$ $mVm^{-1}$ |
| Increasing including maximum (1997 Feb 7-2002 June 27) | 6 | **0.97** | 0.58 | 0.94 | -0.66 | **0.96** | 0.70 | 0.94 | 11 | 0.76 | -0.05 | 0.86 | -0.03 | **0.92** | 0.08 | **0.87** | 74 | 0.83 | -0.26 | 0.77 | 0.28 | **0.83** | -0.33 | **0.81** |
| Decreasing including minimum (2002 June 28-2009 Aug 5) | 7 | 0.88 | -0.82 | 0.92 | 0.75 | **0.998** | -0.40 | **0.997** | 15 | 0.93 | -0.65 | 0.87 | 0.60 | **0.989** | -0.32 | **0.983** | 96 | 0.67 | -0.29 | 0.81 | 0.31 | **0.93** | -0.30 | **0.91** |
| Solar cycle 23 (1997 Feb 7-2009 Aug 5) | 13 | 0.88 | -0.37 | 0.80 | 0.34 | **0.986** | -0.14 | **0.977** | 26 | 0.86 | -0.37 | 0.78 | 0.34 | **0.98** | -0.19 | **0.97** | 170 | 0.74 | -0.26 | 0.73 | 0.27 | **0.90** | -0.30 | **0.88** |
| Extended period 1970-2011 | 42 | 0.85 | 0.02 | 0.70 | -0.03 | **0.97** | 0.07 | **0.96** | 84 | 0.83 | -0.05 | 0.72 | 0.06 | **0.96** | -0.05 | **0.95** | 569 | 0.72 | -0.13 | 0.70 | 0.16 | **0.88** | -0.13 | **0.87** |

| Periods | Correlation coefficients (R) Daily | | | | | | | | 3-Hourly | | | | | | | | Hourly | | | | | | | |
|---|---|---|---|---|---|---|---|---|---|---|---|---|---|---|---|---|---|---|---|---|---|---|---|---|
| | No. of data points | B nT | Bz nT | V $kms^{-1}$ | Ey $mVm^{-1}$ | $BV^2 \cdot 10^{-6}$ $mVs^{-1}$ | $BzV^2 \cdot 10^{-6}$ $mVs^{-1}$ | BV $.10^{-3}$ $mVm^{-1}$ | No. of data points | B nT | Bz nT | V $kms^{-1}$ | Ey $mVm^{-1}$ | $BV^2 \cdot 10^{-6}$ $mVs^{-1}$ | $BzV^2 \cdot 10^{-6}$ $mVs^{-1}$ | BV $.10^{-3}$ $mVm^{-1}$ | No. of data point | B nT | Bz nT | V $kms^{-1}$ | Ey $mVm^{-1}$ | $BV^2 \cdot 10^{-6}$ $mVs^{-1}$ | $BzV^2 \cdot 10^{-6}$ $mVs^{-1}$ | BV $.10^{-3}$ $mVm^{-1}$ |
| Increasing including maximum (1997 Feb 7-2002 June 27) | 1996 | **0.66** | -0.39 | 0.52 | 0.43 | **0.81** | -0.40 | **0.81** | 15988 | 0.61 | -0.41 | 0.43 | 0.44 | **0.69** | -0.44 | **0.71** | --- | -- | -- | --- | -- | -- | -- | -- |
| Decreasing including minimum (2002 June 28-2009 Aug 5) | 2564 | **0.69** | -0.41 | 0.52 | 0.46 | **0.82** | -0.42 | **0.82** | 20512 | 0.61 | -0.40 | 0.46 | 0.43 | **0.72** | -0.40 | **0.73** | --- | -- | -- | --- | -- | -- | -- | -- |
| Solar cycle 23 (1997 Feb 7-2009 Aug 5) | 4564 | 0.67 | -0.39 | 0.51 | 0.44 | **0.814** | -0.40 | **0.82** | 36512 | 0.61 | -0.40 | 0.44 | 0.43 | **0.71** | -0.42 | **0.72** | --- | -- | -- | --- | -- | -- | -- | -- |
| Extended period 1970-2011 | 15340 | 0.63 | -0.33 | 0.53 | 0.35 | **0.79** | -0.35 | **0.78** | 122720 | 0.56 | -0.38 | 0.45 | 0.42 | **0.69** | -0.42 | **0.69** | --- | -- | -- | --- | -- | -- | -- | -- |

**Table 3b**

Linear correlation coefficient (*R*) of *aa* index with interplanetary plasma/field parameters during different phases of solar cycle 23 at different time resolutions, and for extended period (1970-2011). Best (pink) and second best (yellow) values of *R* are highlighted for each set.

Correlation coefficients (R)

Yearly

| Periods | No. of data points | B nT | Bz nT | V $kms^{-1}$ | Ey $mVm^{-1}$ | $BV^2$ $\cdot 10^{-6}$ $mVs^{-1}$ | $BzV^2$ $\cdot 10^{-6}$ $mVs^{-1}$ | BV $.10^{-3}$ $mVm^{-1}$ |
|---|---|---|---|---|---|---|---|---|
| Increasing including maximum (1997 Feb 7-2002 June 27) | 6 | 0.88 | 0.60 | 0.95 | -0.67 | **0.97** | 0.70 | **0.94** |
| Decreasing including minimum (2002 June 28-2009 Aug 5) | 7 | 0.96 | -0.79 | 0.94 | 0.72 | **0.998** | -0.36 | 0.993 |
| Solar cycle 23 (1997 Feb 7-2009 Aug 5) | 13 | 0.86 | -0.36 | 0.83 | 0.33 | **0.99** | -0.12 | **0.97** |
| Extended period 1970-2011 | 42 | 0.82 | -0.01 | 0.75 | -0.007 | **0.98** | 0.05 | **0.95** |

Half yearly

| Periods | No. of data points | B nT | Bz nT | V $kms^{-1}$ | Ey $mVm^{-1}$ | $BV^2$ $\cdot 10^{-6}$ $mVs^{-1}$ | $BzV^2$ $\cdot 10^{-6}$ $mVs^{-1}$ | BV $.10^{-3}$ $mVm^{-1}$ |
|---|---|---|---|---|---|---|---|---|
| Increasing including maximum (1997 Feb 7-2002 June 27) | 11 | 0.76 | 0.03 | 0.82 | -0.05 | **0.93** | 0.09 | **0.88** |
| Decreasing including minimum (2002 June 28-2009 Aug 5) | 15 | 0.92 | -0.64 | 0.85 | 0.56 | **0.993** | -0.26 | **0.98** |
| Solar cycle 23 (1997 Feb 7-2009 Aug 5) | 26 | 0.84 | -0.35 | 0.82 | 0.31 | **0.98** | -0.15 | **0.96** |
| Extended period 1970-2011 | 84 | 0.80 | -0.07 | 0.77 | 0.08 | **0.97** | -0.05 | **0.94** |

27 days

| Periods | No. of data point | B nT | Bz nT | V $kms^{-1}$ | Ey $mVm^{-1}$ | $BV^2$ $\cdot 10^{-6}$ $mVs^{-1}$ | $BzV^2$ $\cdot 10^{-6}$ $mVs^{-1}$ | BV $.10^{-3}$ $mVm^{-1}$ |
|---|---|---|---|---|---|---|---|---|
| Increasing including maximum (1997 Feb 7-2002 June 27) | 74 | 0.71 | -0.26 | 0.79 | 0.27 | **0.86** | -0.33 | **0.84** |
| Decreasing including minimum (2002 June 28-2009 Aug 5) | 96 | 0.83 | -0.27 | 0.85 | 0.26 | **0.95** | -0.28 | **0.93** |
| Solar cycle 23 (1997 Feb 7-2009 Aug 5) | 170 | 0.75 | -0.23 | 0.78 | 0.24 | **0.93** | -0.27 | **0.90** |
| Extended period 1970-2011 | 569 | 0.72 | -0.14 | 0.75 | 0.16 | **0.91** | -0.14 | **0.88** |

| | Correlation coefficients (R) | | | | | | | | | | | | | | | | | | | | | | | |
|---|---|---|---|---|---|---|---|---|---|---|---|---|---|---|---|---|---|---|---|---|---|---|---|---|
| | Daily | | | | | | | | 3-Hourly | | | | | | | | Hourly | | | | | | | |
| Periods | No. of data points | B nT | Bz nT | V $kms^{-1}$ | Ey $mVm^{-1}$ | $BV^2$ $\cdot10^{-6}$ $mVs^{-1}$ | $BzV^2$ $\cdot10^{-6}$ $mVs^{-1}$ | BV $.10^{-3}$ $mVm^{-1}$ | No. of data points | B nT | Bz nT | V $kms^{-1}$ | Ey $mVm^{-1}$ | $BV^2$ $\cdot10^{-6}$ $mVs^{-1}$ | $BzV^2$ $\cdot10^{-6}$ $mVs^{-1}$ | BV $.10^{-3}$ $mVm^{-1}$ | No. of data point | B nT | Bz nT | V $kms^{-1}$ | Ey $mVm^{-1}$ | $BV^2$ $\cdot10^{-6}$ $mVs^{-1}$ | $BzV^2$ $\cdot10^{-6}$ $mVs^{-1}$ | BV $.10^{-3}$ $mVm^{-1}$ |
| Increasing including maximum (1997 Feb 7-2002 June 27) | 1996 | 0.69 | -0.40 | 0.54 | 0.42 | **0.78** | -0.39 | **0.80** | 15988 | 0.60 | -0.40 | 0.45 | 0.42 | **0.65** | -0.40 | **0.68** | -- | -- | -- | -- | -- | -- | -- | -- |
| Decreasing including minimum (2002 June 28-2009 Aug 5) | 2564 | **0.68** | -0.34 | 0.58 | 0.38 | **0.82** | -0.33 | **0.82** | 20512 | 0.60 | -0.35 | 0.49 | 0.37 | **0.70** | -0.34 | **0.71** | -- | -- | -- | -- | -- | -- | -- | -- |
| Solar cycle 23 (1997 Feb 7-2009 Aug 5) | 4564 | 0.67 | -0.35 | 0.55 | 0.39 | **0.80** | -0.35 | **0.81** | 36511 | 0.59 | -0.37 | 0.47 | 0.39 | **0.68** | -0.37 | **0.70** | -- | -- | -- | -- | -- | -- | -- | -- |
| Extended period 1970-2011 | 15340 | 0.64 | -0.32 | 0.57 | 0.34 | **0.80** | -0.32 | **0.79** | 122720 | 0.56 | -0.36 | 0.48 | 0.39 | **0.67** | -0.37 | **0.68** | -- | -- | -- | -- | -- | -- | -- | -- |

**Table 3c**
Linear correlation coefficient (*R*) of *AE* index with interplanetary plasma/field parameters during different phases of solar cycle 23 at different time resolutions, and for extended period (1970-2011). Best (pink) and second best (yellow) values of *R* are highlighted for each set.

| | Correlation coefficients (R) | | | | | | | | | | | | | | | | | | | | | | | |
|---|---|---|---|---|---|---|---|---|---|---|---|---|---|---|---|---|---|---|---|---|---|---|---|---|
| | Yearly | | | | | | | | Half yearly | | | | | | | | 27 days | | | | | | | |
| Periods | No. of data points | B nT | Bz nT | V $kms^{-1}$ | Ey $mVm^{-1}$ | $BV^2$ $\cdot10^{-6}$ $mVs^{-1}$ | $BzV^2$ $\cdot10^{-6}$ $mVs^{-1}$ | BV $.10^{-3}$ $mVm^{-1}$ | No. of data points | B nT | Bz nT | V $kms^{-1}$ | Ey $mVm^{-1}$ | $BV^2$ $\cdot10^{-6}$ $mVs^{-1}$ | $BzV^2$ $\cdot10^{-6}$ $mVs^{-1}$ | BV $.10^{-3}$ $mVm^{-1}$ | No. of data point | B nT | Bz nT | V $kms^{-1}$ | Ey $mVm^{-1}$ | $BV^2$ $\cdot10^{-6}$ $mVs^{-1}$ | $BzV^2$ $\cdot10^{-6}$ $mVs^{-1}$ | BV $.10^{-3}$ $mVm^{-1}$ |
| Increasing including maximum (1997 Feb 7-2002 June 27) | 6 | 0.88 | 0.63 | **0.95** | -0.64 | **0.95** | 0.65 | **0.93** | 11 | 0.66 | -0.07 | 0.72 | 0.08 | **0.85** | 0.08 | **0.80** | 74 | 0.48 | -0.31 | **0.77** | 0.33 | **0.73** | -0.36 | 0.67 |
| Decreasing including minimum (2002 June 28-2009 Aug 5) | 7 | 0.98 | -0.84 | 0.92 | 0.77 | **0.994** | -0.44 | **0.999** | 15 | 0.95 | -0.67 | 0.89 | 0.58 | **0.998** | -0.29 | **0.992** | 96 | 0.85 | -0.35 | **0.81** | 0.33 | **0.92** | -0.35 | **0.92** |
| Solar cycle 23 (1997 Feb 7-2009 Aug 5) | 13 | 0.90 | -0.43 | 0.77 | 0.40 | **0.97** | -0.22 | **0.98** | 26 | 0.88 | -0.41 | 0.77 | 0.37 | **0.96** | -0.22 | **0.97** | 170 | 0.74 | -0.30 | **0.72** | 0.30 | **0.87** | -0.33 | **0.87** |
| Extended period 1970-2011 | 42 | 0.84 | -0.11 | 0.73 | 0.09 | **0.96** | -0.04 | **0.95** | 84 | 0.81 | -0.17 | 0.77 | 0.18 | **0.94** | -0.04 | **0.93** | 569 | 0.70 | -0.19 | 0.69 | 0.21 | **0.84** | -0.19 | **0.83** |

| Periods | Correlation coefficients (R) | | | | | | | | | | | | | | | | | | | | | | | |
|---|---|---|---|---|---|---|---|---|---|---|---|---|---|---|---|---|---|---|---|---|---|---|---|---|
| | Daily | | | | | | | | 3-Hourly | | | | | | | | Hourly | | | | | | | |
| | No. of data points | B nT | Bz nT | V $kms^{-1}$ | Ey $mVm^{-1}$ | $BV^2$ $\cdot 10^{-6}$ $mVs^{-1}$ | $BzV^2$ $\cdot 10^{-6}$ $mVs^{-1}$ | BV $.10^{-3}$ $mVm^{-1}$ | No. of data points | B nT | Bz nT | V $kms^{-1}$ | Ey $mVm^{-1}$ | $BV^2$ $\cdot 10^{-6}$ $mVs^{-1}$ | $BzV^2$ $\cdot 10^{-6}$ $mVs^{-1}$ | BV $.10^{-3}$ $mVm^{-1}$ | No. of data point | B nT | Bz nT | V $kms^{-1}$ | Ey $mVm^{-1}$ | $BV^2$ $\cdot 10^{-6}$ $mVs^{-1}$ | $BzV^2$ $\cdot 10^{-6}$ $mVs^{-1}$ | BV $.10^{-3}$ $mVm^{-1}$ |
| Increasing including maximum (1997 Feb 7-2002 June 27) | 1996 | 0.52 | **-0.61** | 0.51 | **0.60** | 0.58 | -0.53 | **0.60** | 15988 | 0.43 | **-0.63** | 0.41 | **0.61** | 0.43 | -0.54 | 0.48 | 47904 | 0.39 | **-0.56** | 0.38 | **0.54** | 0.40 | -0.47 | 0.44 |
| Decreasing including minimum (2002 June 28-2009 Aug 5) | 2564 | 0.63 | -0.49 | 0.63 | 0.58 | **0.74** | -0.45 | **0.76** | 20512 | 0.51 | -0.54 | 0.52 | 0.52 | **0.56** | -0.45 | **0.60** | 61632 | 0.47 | -0.48 | 0.48 | 0.47 | **0.52** | -0.41 | **0.55** |
| Solar cycle 23 (1997 Feb 7-2009 Aug 5) | 4564 | 0.58 | -0.54 | 0.57 | 0.53 | **0.68** | -0.48 | **0.69** | 36512 | 0.47 | **-0.58** | 0.46 | **0.56** | 0.51 | -0.47 | 0.54 | 106512 | 0.43 | **-0.52** | 0.43 | **0.50** | 0.47 | -0.44 | **0.50** |
| Extended period 1970-2011 | 15340 | 0.53 | -0.49 | **0.57** | 0.49 | **0.67** | -0.46 | **0.67** | 122720 | 0.44 | **-0.56** | 0.47 | **0.56** | 0.52 | -0.50 | **0.54** | 368160 | 0.41 | **-0.51** | 0.43 | **0.51** | 0.48 | -0.46 | **0.50** |

**Table 3d**

Linear correlation coefficient (*R*) of *Dst* index with interplanetary plasma/field parameters during different phases of solar cycle 23 at different time resolutions, and for extended period (1970-2011). Best (pink) and second best (yellow) values of *R* are highlighted for each set.

| Periods | Correlation coefficients (R) | | | | | | | | | | | | | | | | | | | | | | | |
|---|---|---|---|---|---|---|---|---|---|---|---|---|---|---|---|---|---|---|---|---|---|---|---|---|
| | Yearly | | | | | | | | Half yearly | | | | | | | | 27 days | | | | | | | |
| | No. of data points | B nT | Bz nT | V $kms^{-1}$ | Ey $mVm^{-1}$ | $BV^2$ $\cdot 10^{-6}$ $mVs^{-1}$ | $BzV^2$ $\cdot 10^{-6}$ $mVs^{-1}$ | BV $.10^{-3}$ $mVm^{-1}$ | No. of data points | B nT | Bz nT | V $kms^{-1}$ | Ey $mVm^{-1}$ | $BV^2$ $\cdot 10^{-6}$ $mVs^{-1}$ | $BzV^2$ $\cdot 10^{-6}$ $mVs^{-1}$ | BV $.10^{-3}$ $mVm^{-1}$ | No. of data point | B nT | Bz nT | V $kms^{-1}$ | Ey $mVm^{-1}$ | $BV^2$ $\cdot 10^{-6}$ $mVs^{-1}$ | $BzV^2$ $\cdot 10^{-6}$ $mVs^{-1}$ | BV $.10^{-3}$ $mVm^{-1}$ |
| Increasing including maximum (1997 Feb 7-2002 June 27) | 6 | **-0.67** | -0.35 | -0.46 | 0.49 | -0.61 | -0.61 | **-0.63** | 11 | -0.40 | 0.35 | **-0.72** | -0.26 | -0.46 | 0.16 | **-0.43** | 74 | -0.54 | 0.43 | -0.41 | -0.42 | **-0.55** | 0.46 | **-0.56** |
| Decreasing including minimum (2002 June 28-2009 Aug 5) | 7 | -0.93 | 0.73 | -0.93 | -0.64 | **-0.97** | 0.31 | **-0.96** | 15 | **-0.88** | 0.55 | **-0.87** | -0.49 | -0.84 | 0.29 | **-0.87** | 96 | **-0.81** | 0.34 | -0.57 | -0.32 | -0.75 | 0.30 | **-0.80** |
| Solar cycle 23 (1997 Feb 7-2009 Aug 5) | 13 | **-0.92** | 0.42 | -0.54 | -0.34 | -0.83 | 0.12 | **-0.90** | 26 | **-0.82** | 0.44 | -0.52 | -0.39 | **-0.77** | 0.24 | **-0.82** | 170 | **-0.68** | 0.39 | -0.44 | -0.37 | -0.66 | 0.39 | **-0.70** |
| Extended period 1970-2011 | 42 | **-0.85** | -0.003 | -0.33 | 0.05 | -0.76 | -0.09 | **-0.83** | 84 | **-0.78** | 0.13 | -0.42 | -0.12 | -0.76 | 0.10 | **-0.80** | 569 | -0.65 | 0.21 | -0.46 | -0.20 | **-0.68** | 0.19 | **-0.71** |

| Periods | Correlation coefficients (R) | | | | | | | | | | | | | | | | | | | | | | | |
|---|---|---|---|---|---|---|---|---|---|---|---|---|---|---|---|---|---|---|---|---|---|---|---|---|
| | Daily | | | | | | | | 3-Hourly | | | | | | | | Hourly | | | | | | | |
| | No. of data points | B nT | Bz nT | V $kms^{-1}$ | Ey $mVm^{-1}$ | $BV^2$ $\cdot10^{-6}$ $mVs^{-1}$ | $BzV^2$ $\cdot10^{-6}$ $mVs^{-1}$ | BV $.10^{-3}$ $mVm^{-1}$ | No. of data points | B nT | Bz nT | V $kms^{-1}$ | Ey $mVm^{-1}$ | $BV^2$ $\cdot10^{-6}$ $mVs^{-1}$ | $BzV^2$ $\cdot10^{-6}$ $mVs^{-1}$ | BV $.10^{-3}$ $mVm^{-1}$ | No. of data point | B nT | Bz nT | V $kms^{-1}$ | Ey $mVm^{-1}$ | $BV^2$ $\cdot10^{-6}$ $mVs^{-1}$ | $BzV^2$ $\cdot10^{-6}$ $mVs^{-1}$ | BV $.10^{-3}$ $mVm^{-1}$ |
| Increasing including maximum (1997 Feb 7-2002 June 27) | 1996 | -0.46 | 0.39 | -0.48 | -0.36 | **-0.61** | 0.29 | **-0.59** | 15988 | -0.40 | 0.32 | **-0.45** | -0.30 | **-0.49** | 0.25 | **-0.49** | 47904 | -0.38 | 0.29 | -0.44 | -0.27 | **-0.48** | 0.22 | **-0.57** |
| Decreasing including minimum (2002 June 28-2009 Aug 5) | 2564 | **-0.51** | 0.45 | **-0.51** | -0.45 | **-0.64** | 0.41 | **-0.64** | 20512 | -0.43 | 0.36 | -0.47 | -0.36 | **-0.56** | 0.31 | **-0.55** | 61632 | -0.42 | 0.31 | -0.47 | -0.31 | **-0.54** | 0.26 | **-0.64** |
| Solar cycle 23 (1997 Feb 7-2009 Aug 5) | 4564 | -0.49 | 0.42 | -0.47 | -0.40 | **-0.62** | 0.35 | **-0.61** | 36512 | -0.42 | 0.34 | -0.44 | -0.33 | **-0.53** | 0.28 | **-0.52** | 106512 | -0.41 | 0.30 | -0.43 | -0.29 | **-0.51** | 0.24 | **-0.60** |
| Extended period 1970-2011 | 15340 | -0.43 | 0.38 | -0.48 | -0.36 | **-0.59** | 0.34 | **-0.58** | 122720 | **-0.6** | 0.33 | -0.44 | -0.33 | **-0.51** | 0.30 | -0.49 | 368160 | -0.35 | 0.29 | -0.44 | -0.29 | **-0.49** | 0.25 | **-0.48** |

**Table 4**
Best-fit linear equation and correlation coefficients between geomagnetic parameters and *BV* (mV m$^{-1}$), $BV^2$ (mV s$^{-1}$) at different time resolutions

| Geomagnetic parameters | Yearly | | 6-monthly | | 27-day | | Daily | | 3-hourly | | Hourly | |
|---|---|---|---|---|---|---|---|---|---|---|---|---|
| | Relation | *R* | Relation | *R* | Relation | *R* | Relation | *R* | Relation | *R* | Relation | *R* |
| *ap* = | $9.88(BV^2.10^{-6})$ - 1.20 | 0.99 | $9.89(BV^2.10^{-6})$ - 1.20 | 0.98 | $10.72(BV^2.10^{-6})$ -1.36 | 0.90 | $10.15(BV^2.\ 10^{-6})$ -1.56 | 0.82 | $8.91(BV^2.10^{-6})$ -0.17 | 0.71 | -- | -- |
| | $6.02(BV.10^{-3})$ - 4.88 | 0.98 | $6.06(BV.10^{-3})$ - 4.98 | 0.97 | $6.40(BV.10^{-3})$ - 5.62 | 0.88 | $6.96(BV.10^{-3})$ - 7.52 | 0.82 | $6.59(BV.10^{-3})$ -6.54 | 0.72 | | |
| *aa* = | $15.15(BV^2.10^{-6})$ +1.00 | 0.99 | $15.12(BV^2.10^{-6})$ +1.04 | 0.98 | $15.97(BV^2.10^{-6})$ +1.24 | 0.93 | $13.62(BV^2.10^{-6})$ +2.98 | 0.80 | $11.58(BV^2.10^{-6})$ +5.45 | 0.68 | -- | -- |
| | $9.13(BV.10^{-3})$ - 4.36 | 0.97 | $9.18(BV.10^{-3})$ - 4.50 | 0.96 | $9.46(BV.10^{-3})$ - 4.91 | `0.90 | $9.41(BV.10^{-3})$ - 5.19 | 0.81 | $8.69(BV.10^{-3})$ -3.17 | 0.70 | | |
| *AE* = | $140.12(BV^2.10^{-6})$ +10.49 | 0.97 | $140.0(BV^2.10^{-6})$ +10.60 | 0.96 | $131.8(BV^2.10^{-6})$ +33.1 | 0.87 | $88.21(BV^2.\ 10^{-6})$ +79.80 | 0.68 | $68.20(BV^2.10^{-6})$ +105.0 | 0.51 | $67.43(BV^2.10^{-6})$ +105.88 | 0.47 |
| | $89.9(BV.10^{-3})$ - 45.86 | 0.98 | $87.4(BV.10^{-3})$ - 47.3 | 0.97 | $81.3(BV.10^{-3})$ - 27.0 | 0.87 | $62.14(BV.10^{-3})$ + 23.6 | 0.69 | $53.04(BV.10^{-3})$ + 48.81 | 0.54 | $52.26(BV.10^{-3})$ + 50.47 | 0.50 |
| *Dst* = | $-10.26(BV^2.10^{-6})$ – 0.88 | -0.83 | $-10.76(BV^2.10^{-6})$ – 0.23 | -0.77 | $-14.48(BV^2.10^{-6})$ +3.01 | -0.66 | $-10.99(BV^2.\ 10^{-6})$ – 0.37 | -0.62 | $-8.21(BV^2.10^{-6})$ – 3.87 | -0.53 | $-7.97(BV^2.\ 10^{-6})$ - 4.16 | -0.51 |
| | $-6.86(BV.10^{-3})$ + 4.59 | -0.90 | $-7.17(BV.10^{-3})$ + 5.45 | -0.82 | $-9.38(BV.10^{-3})$ + 10.8 | -0.70 | $-7.49(BV.10^{-3})$ - 5.93 | -0.61 | $-5.92(BV.10^{-3})$ + 1.58 | -0.51 | $-5.76(BV.10^{-3})$ + 1.22 | -0.51 |

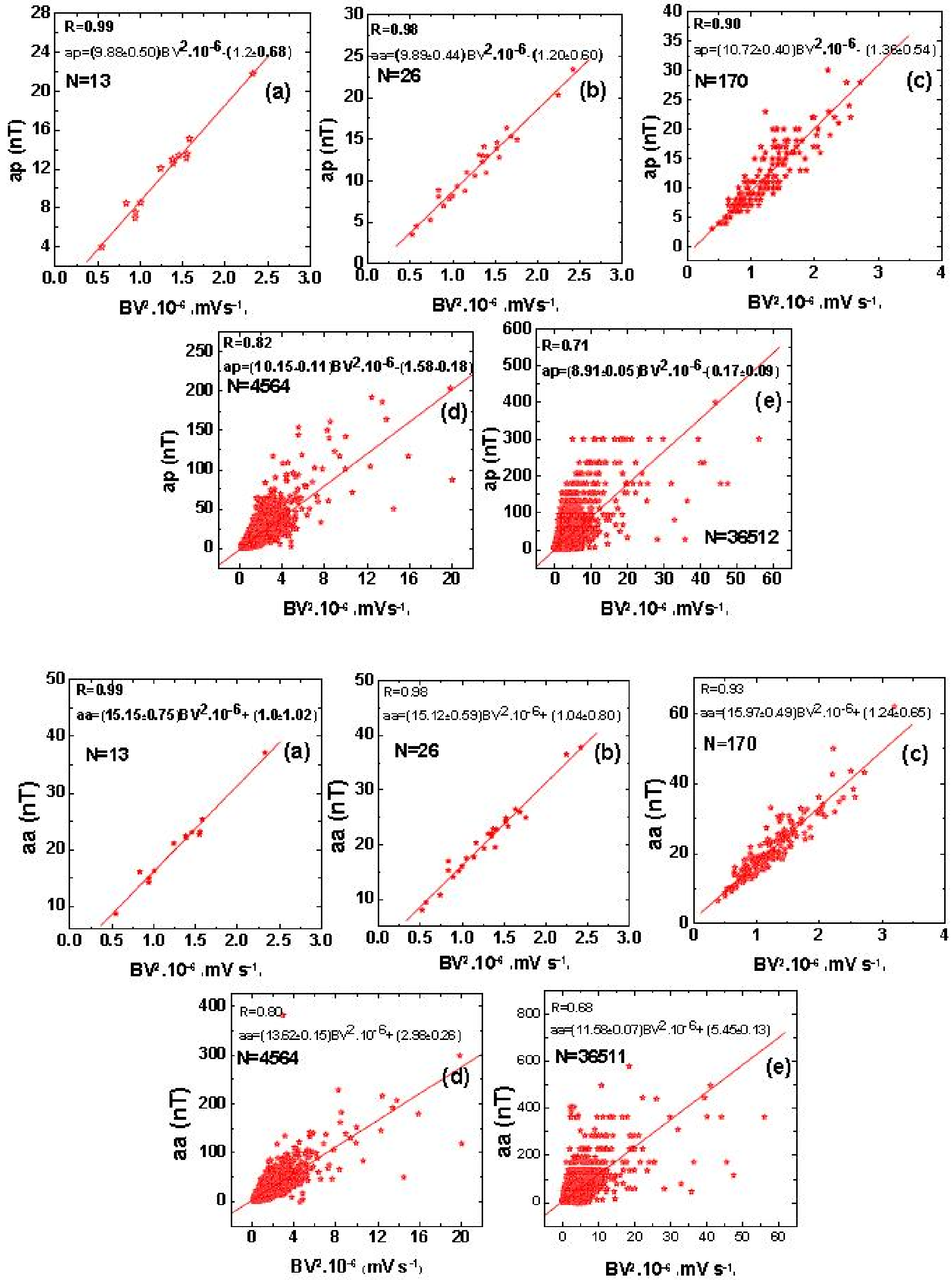

**Fig. 9.** (A) Scatter plot and best-fit linear curve between geomagnetic index ap and $BV^2$ at different time resolutions; (a) yearly, (b) half-yearly, (c) 27-day, (d) daily and (e) 3-hourly. (B) Scatter plot and best fit linear curve between geomagnetic index aa and $BV^2$ at different time resolutions; (a) yearly, (b) half-yearly, (c) 27-day, (d) daily and (e) 3-hourly. (C) Scatter plot and best-fit linear curve between geomagnetic index AE and $BV^2$ at different time resolutions; (a) yearly, (b) half-yearly, (c) 27-day, (d) daily, (e) 3-hourly and (f) hourly. (D) Scatter plot and best-fit linear curve between geomagnetic index Dst and $BV^2$ at different time resolutions; (a) yearly, (b) half-yearly, (c) 27-day, (d) daily,(e) 3-hourly and (f) hourly.

Using 27-day average data (1964-1999), Papitashvili et al. (2000) have revealed that over long time scales, the total interplanetary electric field ($E = BV$), in which the magnetic sphere is immersed, plays a significant role in driving global geomagnetic activity. Sabbah (2000) used yearly average geomagnetic and interplanetary data for the period 1978-1995 and reached at the conclusion that the product $BV$ directly modulates the geomagnetic activity and that this product ($BV$) is more important for the geomagnetic activity modulation rather than the interplanetary magnetic field alone.

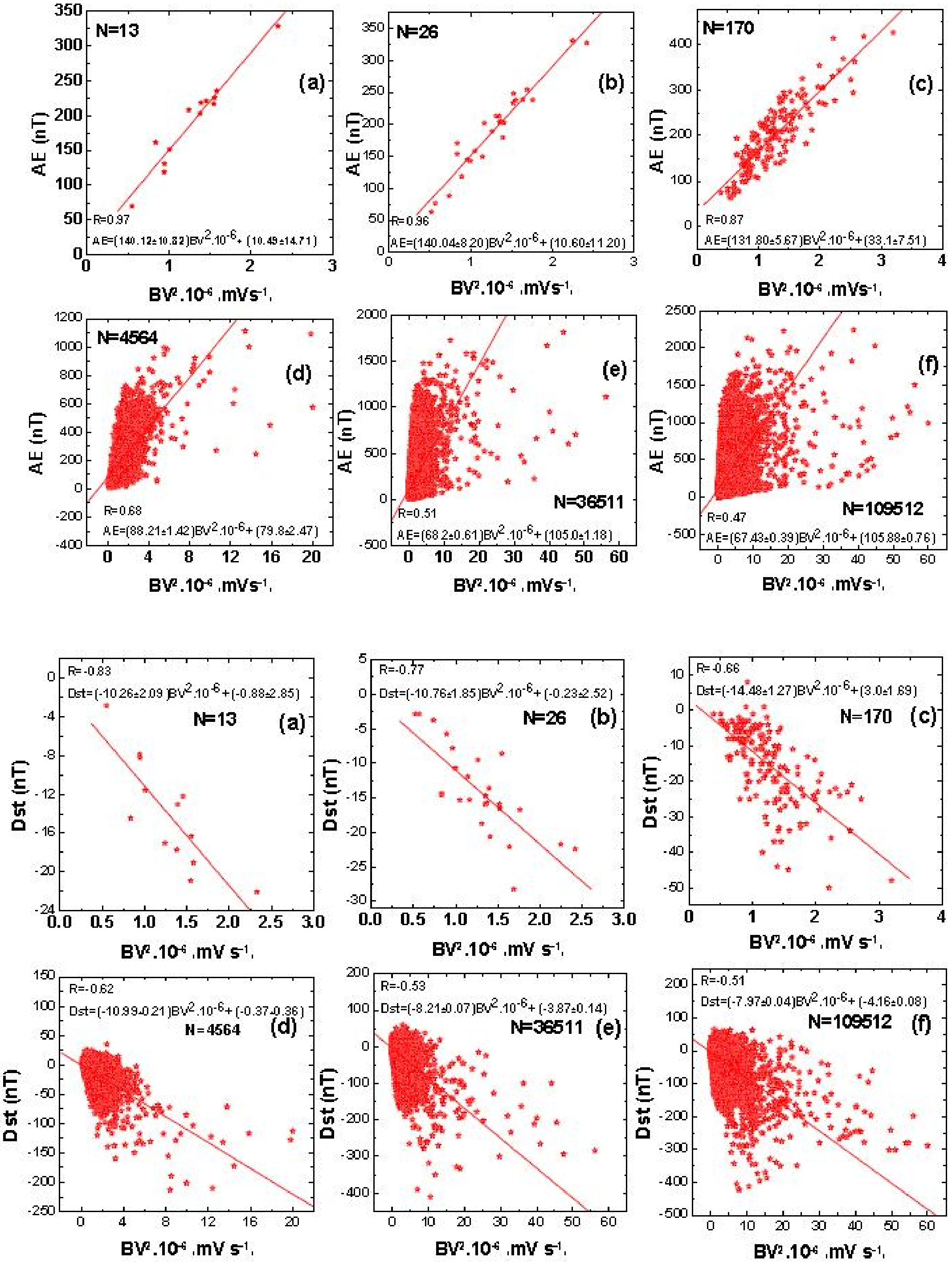

**Fig. 9.** (continued)

Two parameters that appears most in relation with geomagnetic indices are –*B*z (*B*s) and *V* (their product *B*s*V* also). However, there is consider scatter in –*Dst*(min) at low –*B*z values (< 20 nT). Even at considerably large –*B*z and *E*y (-*B*z*V*), -*Dst*(min) range is considerably large (see Kane, 2010; Singh and Badruddin, 2012). It is very likely that some interplanetary parameter(s)/conditions may play a role in enhancing the geomagnetic activity under similar –*B*z and/or *E*y conditions. We suggest that fluctuations in interplanetary electric potential in space [interplanetary electric field *BV*/1000 (mV $m^{-1}$)] and/or time [$BV^2$ (mV $s^{-1}$)] are likely to influence the reconnection rate at the magnetopause to a certain extent when interplanetary field orientation is favourable for reconnection. We suspect that faster variation (spacial and/or time) in interplanetary electric potential might enhance the reconnection rate and increase the amount of energy transfer from the solar wind into the magnetosphere.

## 3. Conclusions

The spacial and/or time variability of interplanetary electric potential (*BV*, $BV^2$) at the magnetopause appears to be an important parameter for solar wind-magnetosphere coupling. When one or both of these variabilities in interplanetary potential are sufficiently large, it is likely to increase the reconnection rate between the solar wind and terrestrial magnetosphere, significantly increasing the geoefficiency of the solar wind.

It is thought that the principal manifestation of geomagnetic storms, measured by *Dst*, is the increase in the ring current intensity, which depends upon the reconnection rate (between the solar wind and the magnetosphere) that allows the solar wind energy transfer into the Earth's magnetosphere/magneto-tail. It is generally believed that a persistent southward IMF (-*B*z) produces increased geomagnetic activity, and dawn-dusk electric field (-*V*x*B*z) is a crucial storm parameter. We suspect that, although the duskward electric field is responsible for initiating the geomagnetic disturbances, the enhanced spacial and/or time fluctuations in the interplanetary electric potential at near the magnetopause is/are the likely additional effect(s) that lead to enhanced coupling between the solar wind and the terrestrial magnetosphere, significantly increasing the geoeffectiveness of the solar wind. However, exact role and mechanism that involves these parameters needs to be identified. We put forward the idea that, when the interplanetary potential at/near the magnetopause fluctuates, in space and/or time, at a sufficiently fast rate, under southward *B*z conditions, it appears to enhance the reconnection rate at the magnetopause. This in turn may cause faster magnetic merging, that occur leading to energy transfer at an enhanced rate from the solar wind to the magnetosphere. However, this hypothesis needs to be tested with high time resolution in situ measurements of interplanetary data. If confirmed, our hypothesis has important implications for solar-terrestrial physics and space weather forecast.

**Acknowledgements**

We gratefully acknowledge the availability of solar, geomagnetic and solar wind plasma/field data through the NASA/GSFC OMNI Web interface and National Geophysical Data Center. We also thank the reviewer, whose comments and suggestions helped us to improve the paper.